\documentclass[12pt,english]{article}

\usepackage[english]{babel}
\usepackage[utf8]{inputenc}
\usepackage{graphicx,xcolor}
\usepackage{epsf}
\usepackage{amssymb}
\usepackage{hyperref}
\usepackage{booktabs}
\usepackage{siunitx}
\usepackage{hyperref}
\usepackage{listings} 
\newcommand{\beq}{\begin{equation}}
\newcommand{\eeq}{\end{equation}}
\newcommand{\bea}{\begin{eqnarray}}
\newcommand{\eea}{\end{eqnarray}}

\newcommand{\as}{\alpha_s\left(\mu_R\right)}
\newcommand{\asq}{\alpha_s^2\left(\mu_R\right)}
\newcommand{\asb}{\bar{\alpha}_s\left(\mu_R\right)}
\newcommand{\cnn}{\chi\left(n,\nu\right)}

\newcommand{\stringa}{\ttfamily\lstinline}
\def\cod#1{{\stringa!#1!}}

\begin{document}
\begin{titlepage}
\begin{center}
{\LARGE \bf Mueller-Navelet jets at LHC: 

BFKL versus high-energy DGLAP}
\end{center}

\vskip 0.5cm

\centerline{F.G.~Celiberto$^{1\ast}$, D.Yu.~Ivanov$^{2,3\P}$, B.~Murdaca$^{1\dag}$
and A.~Papa$^{1\ddagger}$}

\vskip .6cm

\centerline{${}^1$ {\sl Dipartimento di Fisica, Universit\`a della Calabria,}}
\centerline{\sl and Istituto Nazionale di Fisica Nucleare, Gruppo collegato di
Cosenza,}
\centerline{\sl Arcavacata di Rende, I-87036 Cosenza, Italy}

\vskip .2cm

\centerline{${}^2$ {\sl Sobolev Institute of Mathematics, RU-630090 
Novosibirsk, Russia}}

\vskip .2cm

\centerline{${}^3$ {\sl Novosibirsk State University, RU-630090 Novosibirsk, 
Russia}}

\vskip 2cm

\begin{abstract}
The production of forward jets separated by a large rapidity gap at LHC,
the so-called Mueller-Navelet jets, is a fundamental testfield for
perturbative QCD in the high-energy limit. Several analyses have already
provided with evidence about the compatibility of theoretical predictions, 
based on collinear factorization and BFKL resummation of energy logarithms
in the next-to-leading approximation, with the CMS experimental data 
at 7~TeV of center-of-mass energy. However, the question if the same
data can be described also by fixed-order perturbative approaches has not been
yet fully answered. In this paper we provide numerical evidence that
the mere use of partially asymmetric cuts in the transverse momenta 
of the detected jets allows a clear separation between BFKL-resummed and 
fixed-order predictions in some observables related with the Mueller-Navelet 
jet production process.
\end{abstract}


$
\begin{array}{ll}
^{\ast}\mbox{{\it e-mail address:}} &
\mbox{francescogiovanni.celiberto@fis.unical.it}\\
^{\P}\mbox{{\it e-mail address:}} &
\mbox{d-ivanov@math.nsc.ru}\\
^{\dag}\mbox{{\it e-mail address:}} &
\mbox{beatrice.murdaca@fis.unical.it}\\
^{\ddagger}\mbox{{\it e-mail address:}} &
\mbox{alessandro.papa@fis.unical.it}\\
\end{array}
$

\end{titlepage}

\vfill \eject

\section{Introduction}
\label{intro}

It is widely believed now that the inclusive hadroproduction of two jets 
featuring transverse momenta of the same order and much larger than the 
typical hadronic masses and being separated by a large rapidity gap $Y$, 
the so-called Mueller-Navelet jets~\cite{Mueller:1986ey}, is a fundamental 
testfield for perturbative QCD in the high-energy limit, the jet transverse 
momenta providing with the hard scales of the process.

\begin{figure}[tb]
\centering
\includegraphics[scale=0.7]{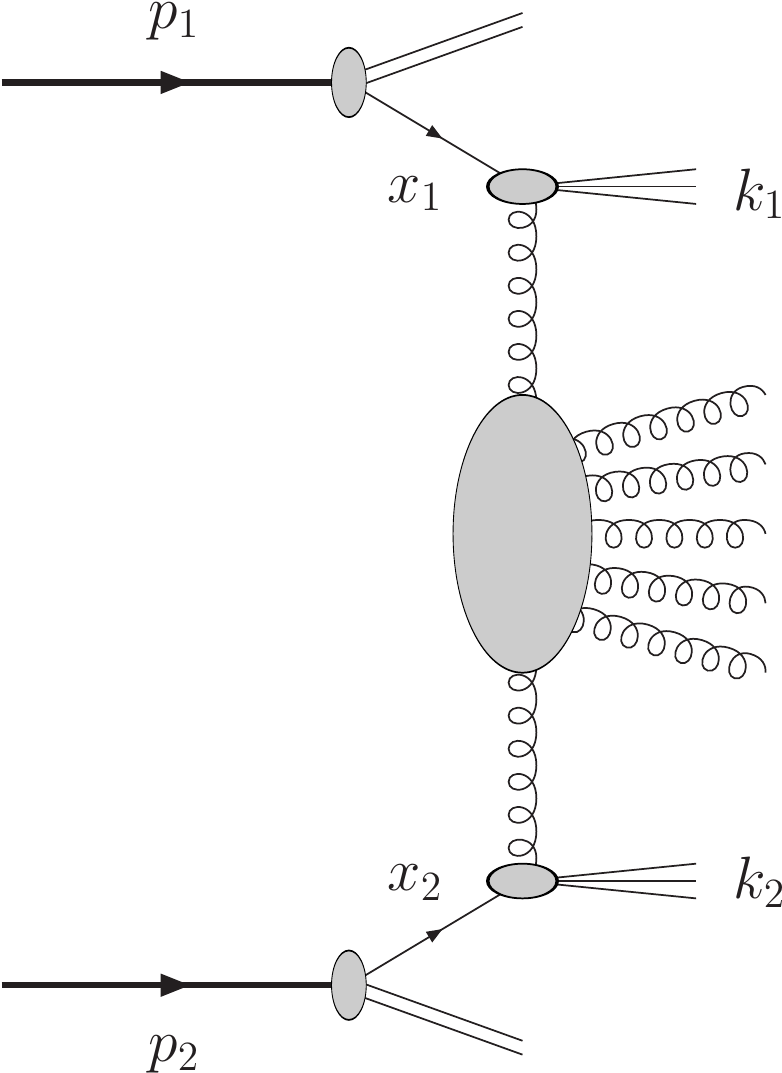}
\caption[]{
Mueller-Navelet jet production process.}
\label{fig:MN}
\end{figure}

At the LHC energies, the theoretical description of this process lies
at the crossing point of two distinct approaches: collinear factorization 
and BFKL~\cite{BFKL} resummation. On one side, at leading twist the
process can be seen as the hard scattering of two partons, each emitted by one
of the colliding hadrons according to the appropriate parton distribution 
function (PDF), see Fig.~\ref{fig:MN}. Collinear factorization takes care
to systematically resum the logarithms of the hard scale, through the 
standard DGLAP evolution~\cite{DGLAP} of the PDFs and the fixed-order 
radiative corrections to the parton scattering cross section.

The other resummation mechanism at work, justified by the large center-of-mass
energy $\sqrt{s}$ available at LHC, is the BFKL resummation of energy 
logarithms, which are so large to compensate the small QCD coupling and must 
therefore be accounted for to all orders of perturbation. These energy 
logarithms are related with the emission of undetected partons between the two 
jets (the larges $s$, the larger the number of partons), which lead to
a reduced azimuthal correlation between the two detected forward jets,
in comparison to the fixed-order DGLAP calculation, where jets are emitted
almost back-to-back.

In the BFKL approach energy logarithms are systematically resummed in the
leading logarithmic approximation (LLA), which means all terms 
$(\alpha_s\ln(s))^n$, and in the next-to-leading logarithmic approximation 
(NLA), which means resummation of all terms $\alpha_s(\alpha_s\ln(s))^n$.
The  process-independent part of such resummation is encoded in the BFKL
Green's function, obeying an iterative integral equation, whose
kernel is known at the next-to-leading order (NLO) both for forward
scattering ({\it i.e.} for $t=0$ and color singlet in the
$t$-channel)~\cite{FL98,CC98} and for any fixed (not growing with energy)
momentum transfer $t$ and any possible two-gluon color state in the
$t$-channel~\cite{Fadin:1998jv,FG00,FF05}.

To get the cross section for Mueller-Navelet jet production and other
related observables, the BFKL Green's function must be convoluted with
two impact factors for the transition from the colliding parton to the
forward jet (the so-called ``jet vertices''). They were first calculated 
with NLO accuracy in~\cite{Bartels:2002yj} and the result was later 
confirmed in~\cite{Caporale:2011cc}. A simpler expression, more 
practical for numerical purposes, was obtained in~\cite{Ivanov:2012ms}
adopting the so-called ``small-cone'' approximation 
(SCA)~\cite{Furman:1981kf,Aversa}, {\it i.e.} for small jet cone
aperture in the rapidity-azimuthal angle plane. A critical comparison
between the latter result and the exact jet vertex in the cases of  
$k_t$ and cone algorithms and their ``small-cone'' versions has been recently 
carried out in~\cite{Colferai:2015zfa}. We stress that, within the 
NLO accuracy, the jet can be formed by either one or two particles and no
more, so that the argument given in~\cite{Colferai:2015zfa} about the 
non-infrared-safety of the all-order extension of the jet algorithm used to
obtain the SCA jet vertex in~\cite{Ivanov:2012ms} does not apply here.

The BFKL approach brings along some extra-sources of systematic uncertainties
with respect to the fixed-order, DGLAP calculation. First of all, in addition
to the renormalization and factorization scales, $\mu_R$ and $\mu_F$, which
appear also in DGLAP, there is a third, artificial normalization scale, 
usually called $s_0$, which must be suitably fixed. Moreover, there is 
compelling evidence that choosing for these scales the values dictated by the
kinematics of the process is not necessarily the best choice when the BFKL
resummation is at work. It is well known, indeed, that the NLO BFKL corrections
for the $n=0$ conformal spin are with opposite sign with respect to the
leading order (LO) result and large in absolute value. This happens
both to the NLO BFKL kernel and to the process-dependent NLO impact factors 
(see, {\it e.g.} Ref.~\cite{Ivanov2006}, for the case of the vector meson 
photoproduction). This calls for some optimization procedure, which can
consist in (i) including some pieces of the (unknown) next-to-NLO corrections,
such as those dictated by renormalization group, as in \emph{collinear
improvement}~\cite{collinear}, or by energy-momentum 
conservation~\cite{Kwiecinski:1999yx}, and/or (ii) suitably choosing the values 
of the energy and renormalization scales, which, though arbitrary within the 
NLO, can have a sizeable numerical impact through subleading terms. Common
optimization methods are those inspired by the \emph{principle of minimum 
sensitivity} (PMS)~\cite{PMS}, the \emph{fast apparent convergence} 
(FAC)~\cite{FAC} and the \emph{Brodsky-LePage-Mackenzie method} 
(BLM)~\cite{BLM}.

This variety of options reflects in the large number of numerical studies
of the Mueller-Navelet jet production process at LHC, both at a center-of-mass
energy of 14~TeV~\cite{Colferai2010,Caporale2013,Salas2013} and  
7~TeV~\cite{Ducloue2013,Ducloue2014,Ducloue:2014koa,Caporale:2014gpa}.
All these studies were concerned with the behavior on $Y$ of azimuthal angle 
correlations between the two measured jets, {\it i.e.} average values of 
$\cos{(n \phi)}$, where $n$ is an integer and $\phi$ is the angle in the
azimuthal plane between the direction of one jet and the opposite direction
of the other jet, and ratios of two such cosines~\cite{sabioV}.  
In particular, one of these analyses~\cite{Ducloue2014}, based on the use of 
a collinearly-improved and energy scales optimized \emph{\`a la} BLM,
found a nice agreement with CMS data~\cite{CMS}.

In a recent paper~\cite{Caporale:2014gpa}, some of us stressed that another
important source of systematics on should be aware of is the 
``representation uncertainty'', deriving from the freedom to use different 
representation of the BFKL cross section, equivalent within the 
NLO~\footnote{We remark that the impact of this uncertainty on azimuthal 
correlations and their ratios is much larger than the one resulting from the 
adoption of different jet algorithms at the NLO.}.
The good agreement between CMS data and BLM-optimized theoretical predictions
gives us perhaps a hint towards the right direction. In the same paper, a list
of issues was presented which, if considered in the experimental analysis,
could help the matching between the way Mueller-Navelet are defined in the
theory and in the experiment. One of these issue was, for instance, the
very measurement of the Mueller-Navelet total cross section, $C_0$, which,
on the theory side, is strongly sensitive both to the representation of
the BFKL amplitude and to the optimization procedure.

In this paper we want to further discuss and expand another of the issues 
listed in the ``Discussion'' section of~\cite{Caporale:2014gpa}, related with
the choice of the experimental cuts in the values of the forward jet transverse
momenta. Since the Born contribution to the cross section $C_0$ is present
only for back-to-back jets, its effect is maximized when \emph{symmetric}
cuts are used; on the contrary, in the case of \emph{asymmetric} cuts,
the Born term is suppressed and the effects of the additional undetected 
hard gluon radiation is enhanced, thus making more visible the BFKL
resummation, in comparison to descriptions based on the fixed-order DGLAP 
approach, in all observables involving $C_0$.

For this purpose, we compare predictions for several azimuthal correlations 
and their ratios obtained, on one side, by a fixed-order DGLAP calculation
at the NLO and, on the other side, by BFKL resummation in the NLA. 

To avoid misunderstanding we note that in what follows our implementation of 
the NLO DGLAP calculation will be an approximate one. We just use here 
NLA BFKL expressions for the observables that are truncated to the 
${\cal O}\left(\alpha_s^3\right)$ order.
In this way we take into account the leading power asymptotic of the exact NLO 
DGLAP prediction and neglect terms that are suppressed by the inverse powers of 
the energy of the parton-parton collisions. Such approach is legitimate in the 
region of large $Y$ which we consider here. The exact implementation of NLO 
DGLAP for Mueller-Navelet jets is important, because it allows to understand 
better the region of applicability of our approach, but it requires more 
involved Monte Carlo calculations (some first results were reported recently 
in~\cite{Colferai-WS}).

To single out the only effect of transverse momentum cuts, we consider just one 
representation of the Mueller-Navelet cross section (the \emph{exponentiated} 
one) and just one optimization procedure (the BLM one, in the two variants 
discussed in~\cite{BLMpaper}).

The paper is organized as follows: in the next section we give the
kinematics and the basic formulae for the Mueller-Navelet jet process
cross section; in section~\ref{results} we present our results;
finally, in section~\ref{conclusions} we draw our conclusions.

\section{Theoretical setup}
\label{theory}

In this section we briefly recall the kinematics of the process and the 
main formulae, referring the reader to previous 
papers~\cite{Caporale2013,Caporale:2014gpa} for the omitted details.

The process under exam is the production of Mueller-Navelet 
jets~\cite{Mueller:1986ey} in proton-proton collisions
\begin{eqnarray}
\label{process}
p(p_1) + p(p_2) \to {\rm jet}(k_{J_1}) + {\rm jet}(k_{J_2})+ X \;,
\end{eqnarray}
where the two jets are characterized by high transverse momenta,
$\vec k_{J_1}^2\sim \vec k_{J_2}^2\gg \Lambda_{QCD}^2$ and large separation
in rapidity; $p_1$ and $p_2$ are taken as Sudakov vectors satisfying
$p_1^2=p_2^2=0$ and $2\left( p_1 p_2\right)=s$.

In QCD collinear factorization the cross section of the process~(\ref{process})
reads
\beq
\frac{d\sigma}{dx_{J_1}dx_{J_2}d^2k_{J_1}d^2k_{J_2}}
=\sum_{i,j=q,{\bar q},g}\int_0^1 dx_1 \int_0^1 dx_2\ f_i\left(x_1,\mu_F\right)
\ f_j\left(x_2,\mu_F\right)\frac{d{\hat\sigma}_{i,j}\left(x_1x_2s,\mu_F\right)}
{dx_{J_1}dx_{J_2}d^2k_{J_1}d^2k_{J_2}}\;,
\eeq
where the $i, j$ indices specify the parton types (quarks $q = u, d, s, c, b$;
antiquarks $\bar q = \bar u, \bar d, \bar s, \bar c, \bar b$; or gluon $g$),
$f_i\left(x, \mu_F \right)$ denotes the initial proton PDFs; $x_{1,2}$ are
the longitudinal fractions of the partons involved in the hard subprocess,
while $x_{J_{1,2}}$ are the jet longitudinal fractions; $\mu_F$ is the
factorization scale; $d\hat\sigma_{i,j}\left(x_1x_2s, \mu_F \right)$ is
the partonic cross section for the production of jets and
$x_1x_2s\equiv\hat s$ is the squared center-of-mass energy of the
parton-parton collision subprocess (see Fig.~\ref{fig:MN}).

The cross section of the process can be presented as
\beq
\frac{d\sigma}
{dy_{J_1}dy_{J_2}\, d|\vec k_{J_1}| \, d|\vec k_{J_2}|
d\phi_{J_1} d\phi_{J_2}}
=\frac{1}{(2\pi)^2}\left[{\cal C}_0+\sum_{n=1}^\infty  2\cos (n\phi )\,
{\cal C}_n\right]\, ,
\eeq
where $\phi=\phi_{J_1}-\phi_{J_2}-\pi$, while ${\cal C}_0$ gives the total
cross section and the other coefficients ${\cal C}_n$ determine the distribution
of the azimuthal angle of the two jets. In the BFKL approach several
NLA-equivalent expressions can be adopted for ${\cal C}_n$. A large list of 
them and of their features can be found in~\cite{Caporale:2014gpa}. For the
purposes of the present analysis, we concentrate on one representation,
the so-called {\it exponentiated} representation, and use it in combination with
the BLM optimization procedure. We recall that BLM setting means choosing 
the scale $\mu_R$ such that it makes vanish completely the $\beta_0$-dependence of a given observable. As discussed in~\cite{Caporale:2014gpa}, we implement
two variants of the BLM method, dubbed $(a)$ and $(b)$, derived 
in~\cite{BLMpaper}. Moreover, we use a common optimal scale for the 
renormalization scale $\mu_R$ and for the factorization scale $\mu_F$.
In~\cite{Caporale:2014gpa} it was shown that this setup 
allows a nice agreement with CMS data for several azimuthal correlations and 
their ratios in the large $Y$ regime.

Introducing, for the sake of brevity, the definitions
\[
Y=\ln\frac{x_{J_1}x_{J_2}s}{|\vec k_{J_1}||\vec k_{J_2}|}\;,
\;\;\;\;\;
Y_0=\ln\frac{s_0}{|\vec k_{J_1}||\vec k_{J_2}|}\;,
\]
we have then the following expressions for the coefficients ${\cal C}_n$,
in the two variants of BLM setting:
\[
{\cal C}_n^{\rm BFKL_{(a)}}= \frac{x_{J_1}x_{J_2}}{|\vec k_{J_1}|
|\vec k_{J_2}|}\int_{-\infty}^{+\infty}d\nu
\ e^{(Y-Y_0)\left[\bar \alpha_s\left(\mu_R\right)\chi\left(n,\nu\right)
+\bar \alpha_s^2\left(\mu_R\right)
\left( \bar \chi\left(n,\nu\right)-\frac{T^{\beta}}{C_A}\chi\left(n,\nu\right)
-\frac{\beta_0}{8C_A}\chi^2\left(n,\nu\right)\right)\right]}
\]
\beq\label{casea}
\times \alpha_s^2\left(\mu_R\right)
c_1(n,\nu,|\vec k_{J_1}|, x_{J_1}) c_2(n,\nu,|\vec k_{J_2}|,x_{J_2})
\eeq
\[
\times \left[ 1-\frac{2}{\pi} \alpha_s\left(\mu_R\right)T^{\beta}
+ \alpha_s\left(\mu_R\right)\left(\frac{\bar c_1^{\left(1\right)}
(n,\nu,|\vec k_{J_1}|,x_{J_1})}{c_1(n,\nu,|\vec k_{J_1}|, x_{J_1})}
+\frac{\bar c_2^{\left(1\right)}(n,\nu,|\vec k_{J_2}|, x_{J_2})}
{c_2(n,\nu,|\vec k_{J_2}|, x_{J_2})}\right)
\right]\;,
\]
with $\mu_R$ fixed at the value
\beq\label{scalea}
(\mu_R^{\rm BLM})^2=k_{J_1}k_{J_2}\ \exp\left[2\left(1+\frac{2}{3}I\right)
-\frac{5}{3}\right]\;,
\eeq
and
\[
{\cal C}_n^{\rm BFKL_{(b)}}= \frac{x_{J_1}x_{J_2}}{|\vec k_{J_1}|
|\vec k_{J_2}|}\int_{-\infty}^{+\infty}d\nu
\ e^{(Y-Y_0)\left[\bar \alpha_s\left(\mu_R\right)\chi\left(n,\nu\right)
+ \bar \alpha_s^2\left(\mu_R\right)
\left( \bar \chi\left(n,\nu\right)-\frac{T^{\beta}}{C_A}
\chi\left(n,\nu\right)\right)\right]}
\]
\beq\label{caseb}
\times \alpha_s^2\left(\mu_R\right)
c_1(n,\nu,|\vec k_{J_1}|, x_{J_1}) c_2(n,\nu,|\vec k_{J_2}|,x_{J_2})
\eeq
\[
\times \left[1+\alpha_s\left(\mu_R\right)\left(\frac{\beta_0}{4\pi}
\chi\left(n,\nu\right)
- 2\frac{T^{\beta}}{\pi}\right)
+ \alpha_s\left(\mu_R\right)\left(\frac{\bar c_1^{\left(1\right)}
(n,\nu,|\vec k_{J_1}|,x_{J_1})}{c_1(n,\nu,|\vec k_{J_1}|, x_{J_1})}
+\frac{\bar c_2^{\left(1\right)}(n,\nu,|\vec k_{J_2}|, x_{J_2})}
{c_2(n,\nu,|\vec k_{J_2}|, x_{J_2})}\right)
\right]\;,
\]
with $\mu_R$ fixed at the value
\beq\label{scaleb}
(\mu_R^{\rm BLM})^2=k_{J_1}k_{J_2}\ \exp\left[2\left(1+\frac{2}{3}I\right)
-\frac{5}{3}+\frac{1}{2}\chi\left(\nu,n\right)\right]\;.
\eeq
In Eqs.~(\ref{casea}) and~(\ref{caseb}), $\bar \alpha_s(\mu_R) 
\equiv \alpha_s(\mu_R) N_c/\pi$, with $N_c$ the number of colors,
\beq
\beta_0=\frac{11}{3} N_c - \frac{2}{3}n_f
\eeq
is the first coefficient of the QCD $\beta$-function,
\beq
\chi\left(n,\nu\right)=2\psi\left( 1\right)-\psi\left(\frac{n}{2}
+\frac{1}{2}+i\nu \right)-\psi\left(\frac{n}{2}+\frac{1}{2}-i\nu \right)
\eeq
is the LO BFKL characteristic function,
\beq
\label{c1}
c_1(n,\nu,|\vec k|,x)=2\sqrt{\frac{C_F}{C_A}}
(\vec k^{\,2})^{i\nu-1/2}\,\left(\frac{C_A}{C_F}f_g(x,\mu_F)
+\sum_{a=q,\bar q}f_q(x,\mu_F)\right)
\eeq
and
\beq
\label{c2}
c_2(n,\nu,|\vec k|,x)=\biggl[c_1(n,\nu,|\vec k|,x) \biggr]^* \;,
\eeq
are the LO jet vertices in the $\nu$-representation. The remaining objects
are related with the NLO corrections of the BFKL kernel ($\bar \chi(n,\nu)$)
and of the jet vertices in the small-cone approximation
($c_{1,2}^{(1)}(n,\nu,|\vec k_{J_{1,2}}|, x_{J_{1,2}})$)
in the $\nu$-representation. Their expressions are given in Eqs.~(23), (36)
and~(37) of Ref.~\cite{Caporale2013}. Moreover, 
\begin{eqnarray*}
T^{\beta} = -\frac{\beta_0}{2}\left( 1+\frac{2}{3}I \right)\;,\\
\end{eqnarray*}
where $I=-2\int_0^1dx\frac{\ln\left(x\right)}{x^2-x+1}\simeq 2.3439$ and 
$\bar c_{1,2}^{(1)}(n,\nu,|\vec k_{J_2}|, x_{J_2})$ are the same as
$c_{1,2}^{(1)}(n,\nu,|\vec k_{J_{1,2}}|, x_{J_{1,2}})$ with the terms proportional to 
$\beta_0$ removed. The scale $s_0$ entering $Y_0$ is artificial. It is 
introduced in the BFKL approach at the time to perform
the Mellin transform from the $s$-space to the complex angular momentum plane
and cancels in the full expression, up to terms beyond the NLA.
In the following it will always be fixed at the ``natural'' value $Y_0=0$.

In the fixed-order DGLAP approach at the NLO, the expressions for the 
coefficients ${\cal C}_n$ are nothing but the truncation of the BFKL
expressions up to inclusions of NLO terms and read
\[
{\cal C}_n^{\rm DGLAP_{(a)}}  = \frac{x_{J_1}x_{J_2}}{|\vec k_{J_1}|
|\vec k_{J_2}|}\int_{-\infty}^{+\infty}d\nu
\ \asq 
c_1(n,\nu,|\vec k_{J_1}|, x_{J_1}) c_2(n,\nu,|\vec k_{J_2}|,x_{J_2})
\]
\beq
\label{dglap}
\times 
\left[ 1 - \frac{2}{\pi} \as T^{\beta}
+ \asb \left( Y - Y_0 \right) \cnn \right.
\eeq
\[
\left.
+ \as \left(\frac{\bar c_1^{\left(1\right)}
(n,\nu,|\vec k_{J_1}|,x_{J_1})}{c_1(n,\nu,|\vec k_{J_1}|, x_{J_1})}
+\frac{\bar c_2^{\left(1\right)}(n,\nu,|\vec k_{J_2}|, x_{J_2})}
{c_2(n,\nu,|\vec k_{J_2}|, x_{J_2})}\right)\right] \;,
\]
\[
{\cal C}_n^{\rm DGLAP_{(b)}}  = \frac{x_{J_1}x_{J_2}}{|\vec k_{J_1}|
|\vec k_{J_2}|}\int_{-\infty}^{+\infty}d\nu
\ \asq 
c_1(n,\nu,|\vec k_{J_1}|, x_{J_1}) c_2(n,\nu,|\vec k_{J_2}|,x_{J_2})
\]
\beq
\label{dglapb}
\times 
\left[ 1 +\alpha_s\left(\mu_R\right)\left(\frac{\beta_0}{4\pi}
\chi\left(n,\nu\right)
- 2\frac{T^{\beta}}{\pi}\right)
+ \asb \left( Y - Y_0 \right) \cnn \right.
\eeq
\[
\left.
+ \as \left(\frac{\bar c_1^{\left(1\right)}
(n,\nu,|\vec k_{J_1}|,x_{J_1})}{c_1(n,\nu,|\vec k_{J_1}|, x_{J_1})}
+\frac{\bar c_2^{\left(1\right)}(n,\nu,|\vec k_{J_2}|, x_{J_2})}
{c_2(n,\nu,|\vec k_{J_2}|, x_{J_2})}\right)\right] \;,
\]
which we will use in the following with the two possible choices 
$(a)$ and $(b)$ of the optimal scales, given in Eqs.~(\ref{scalea}) 
and~(\ref{scaleb}), respectively. It is worth mentioning that our DGLAP 
expressions, (\ref{dglap}) and~(\ref{dglapb}), do not actually depend on $Y_0$. 
The corresponding terms in the r.h.s. of~(\ref{dglap}) and~(\ref{dglapb}) are 
cancelled by similar terms in $c^{(1)}_{1, 2}$, see~\cite{Caporale2013}.

We note that, in our way to implement the BLM procedure (see~\cite{BLMpaper}),
the final expressions are given in terms of $\alpha_s$ in the 
$\overline{\rm MS}$ scheme, although in one intermediate step the MOM scheme
was used.

\section{Numerical analysis}
\label{results}

\subsection{Results}

In this Section we present our results for the dependence on the
rapidity separation between the detected jets, $Y=y_{J_1}-y_{J_2}$, 
of ratios ${\cal R}_{nm}\equiv{\cal C}_n/{\cal C}_m$ between the 
coefficients ${\cal C}_n$. Among them, the ratios of
the form $R_{n0}$ have a simple physical interpretation, being the azimuthal
correlations $\langle \cos(n\phi)\rangle$.

In order to match the kinematic cuts used by the CMS collaboration, we will
consider the \emph{integrated coefficients} given by
\beq
\label{Cm_int}
C_n=\int_{y_{1,\rm min}}^{y_{1,\rm max}}dy_1
\int_{y_{2,\rm min}}^{y_{2,\rm max}}dy_2\int_{k_{J_1,\rm min}}^{\infty}dk_{J_1}
\int_{k_{J_2,\rm min}}^{\infty}dk_{J_2} \delta\left(y_1-y_2-Y\right){\cal C}_n
\left(y_{J_1},y_{J_2},k_{J_1},k_{J_2} \right)
\eeq
and their ratios $R_{nm}\equiv C_n/C_m$. We will take jet rapidities in the
range delimited by $y_{1,\rm min}=y_{2,\rm min}=-4.7$  and 
$y_{1,\rm max}=y_{2,\rm max}=4.7$~\footnote{In~\cite{Caporale:2014gpa} it was 
mistakenly written $y_{i,\rm min}=0$, although all numerical results presented
there were obtained using the correct value for $y_{i,\rm min}$.} 
and consider $Y=3$, 6 and 9. 
As for the jet transverse momenta, differently
from all previous analyses, we make two \emph{asymmetric} choices: 
(1) $k_{J_1,\rm min} = 35$ GeV, $k_{J_2,\rm min} = 45$ GeV and 
(2) $k_{J_1,\rm min} = 35$ GeV, $k_{J_2,\rm min} = 50$ GeV.
The jet cone size $R$ entering the NLO-jet vertices is fixed at the value 
$R=0.5$, the center-of-mass energy at $\sqrt s=7$ TeV and, as anticipated,
$Y_0=0$. We use the PDF set MSTW2008nlo~\cite{PDF} and the two-loop
running coupling with $\alpha_s\left(M_Z\right)=0.11707$.

\begin{figure}[p]
\centering
   \includegraphics[scale=0.35]{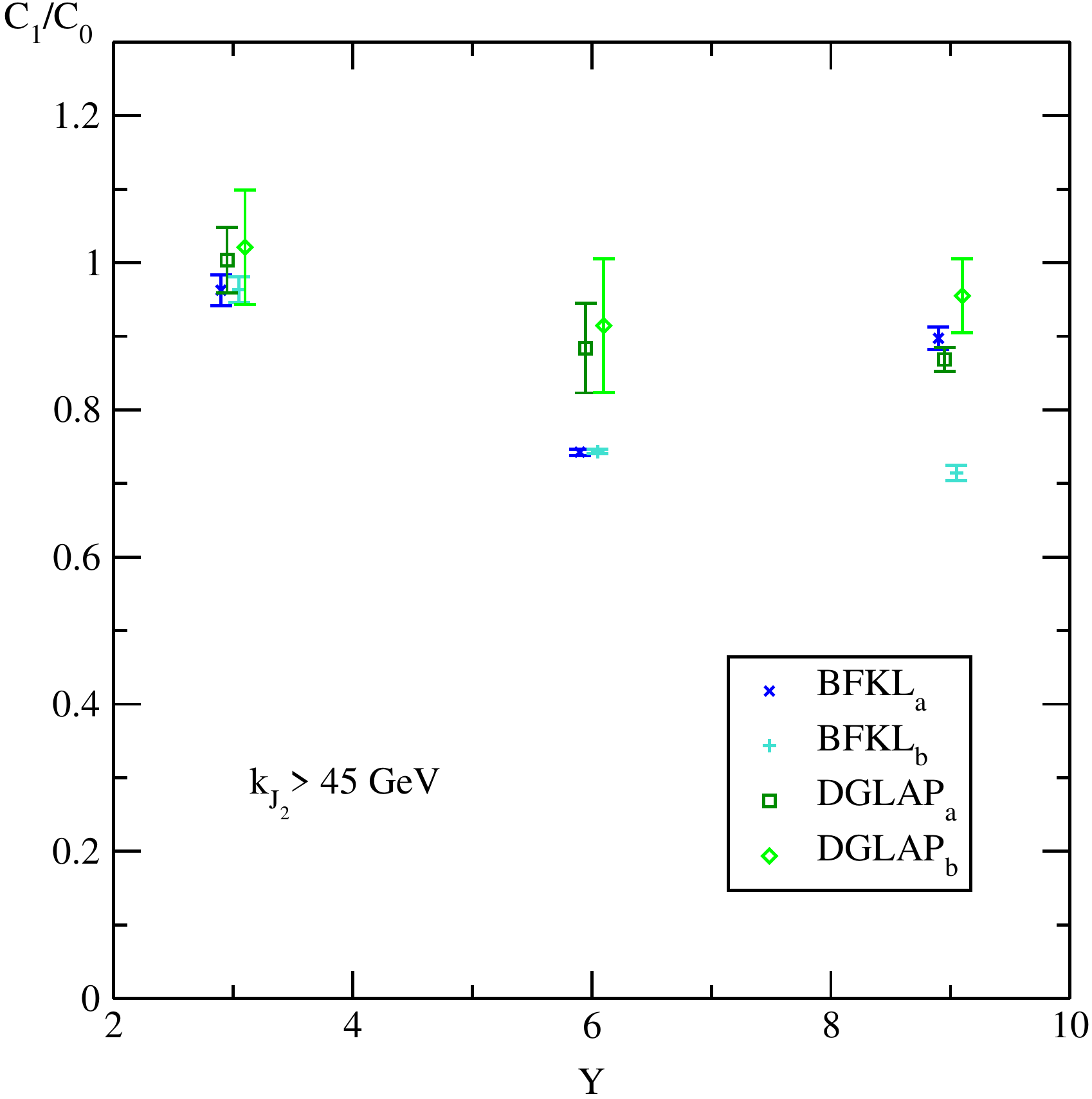}
   \includegraphics[scale=0.35]{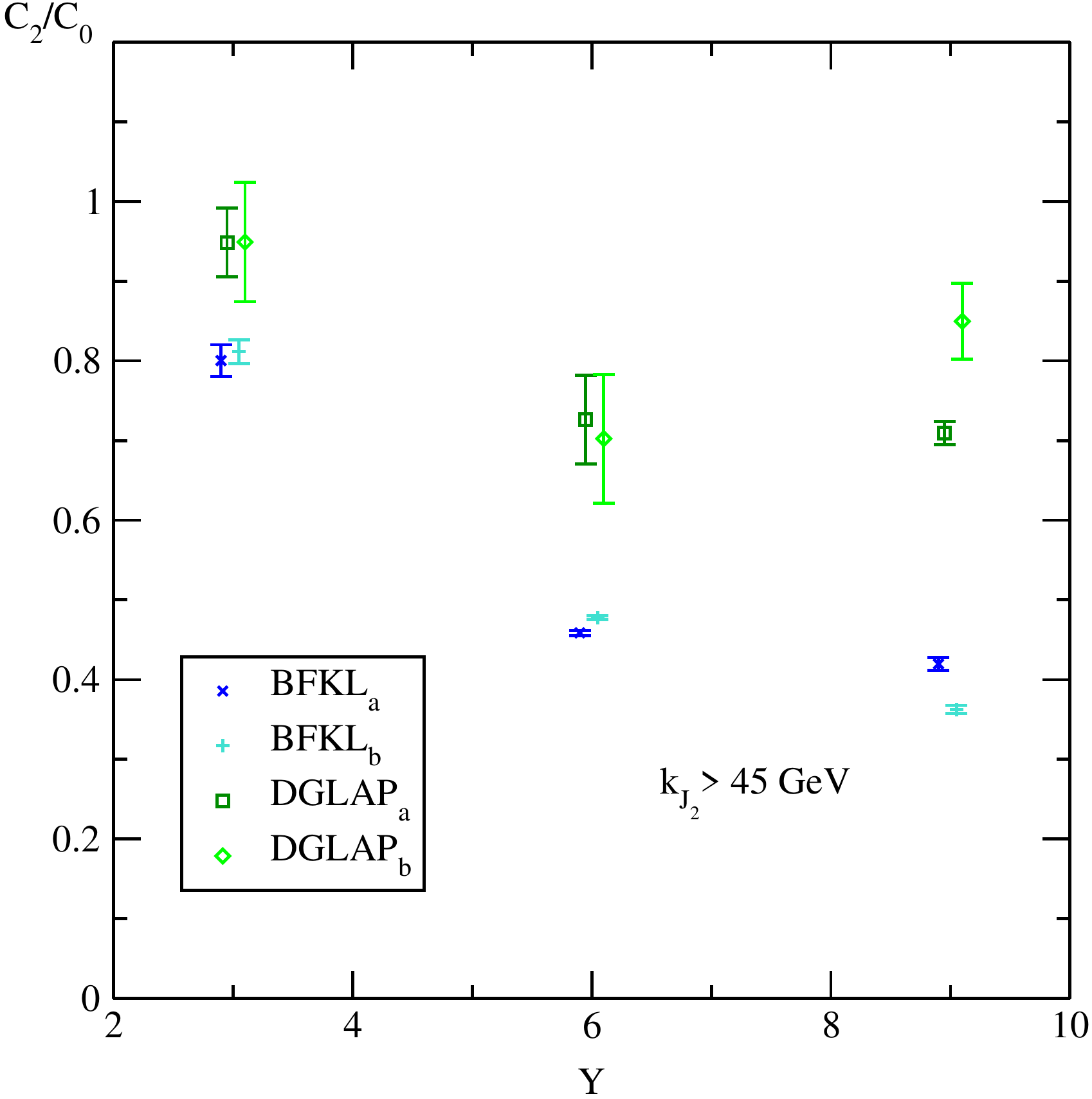}

   \includegraphics[scale=0.35]{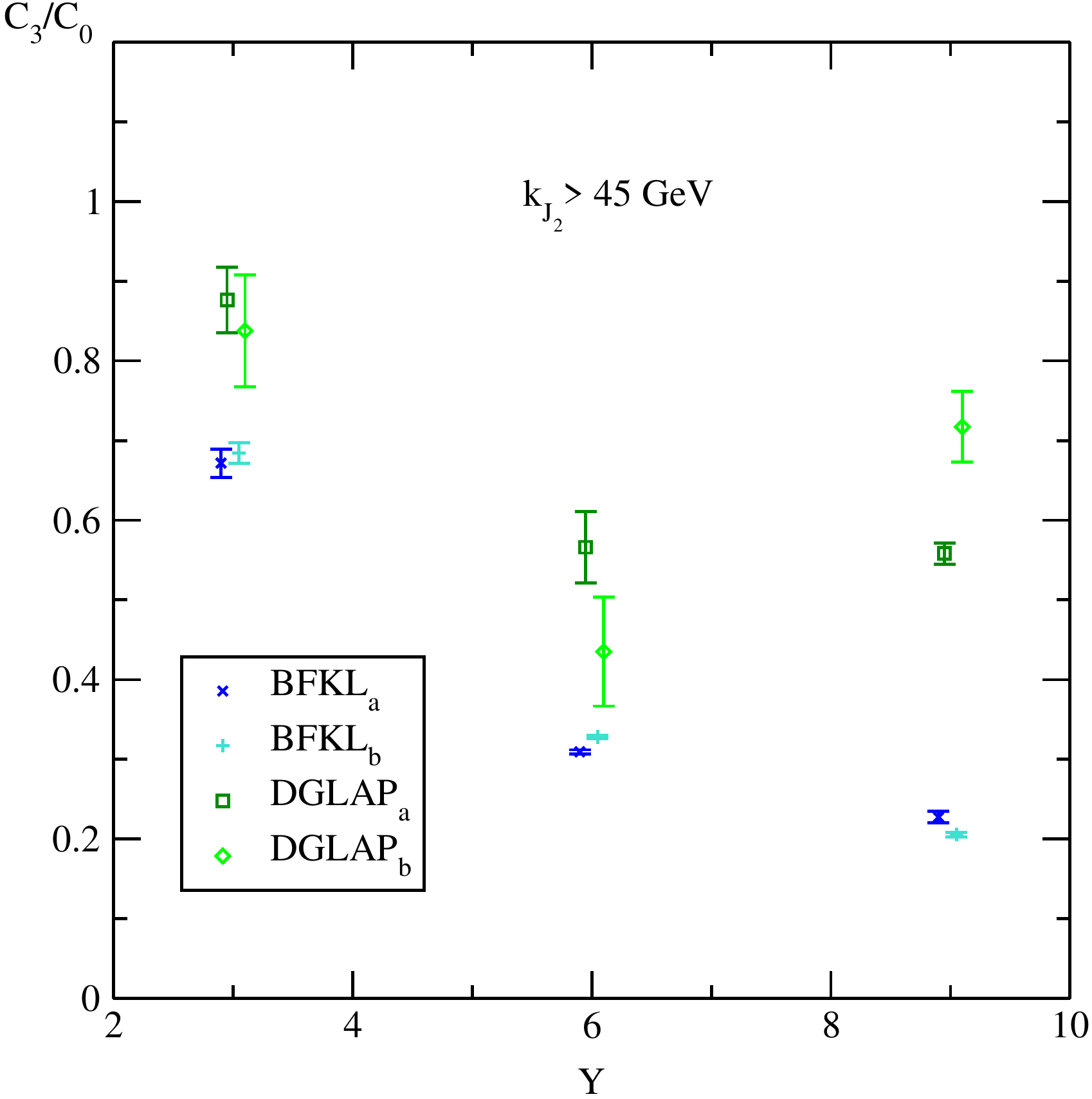}
   \includegraphics[scale=0.35]{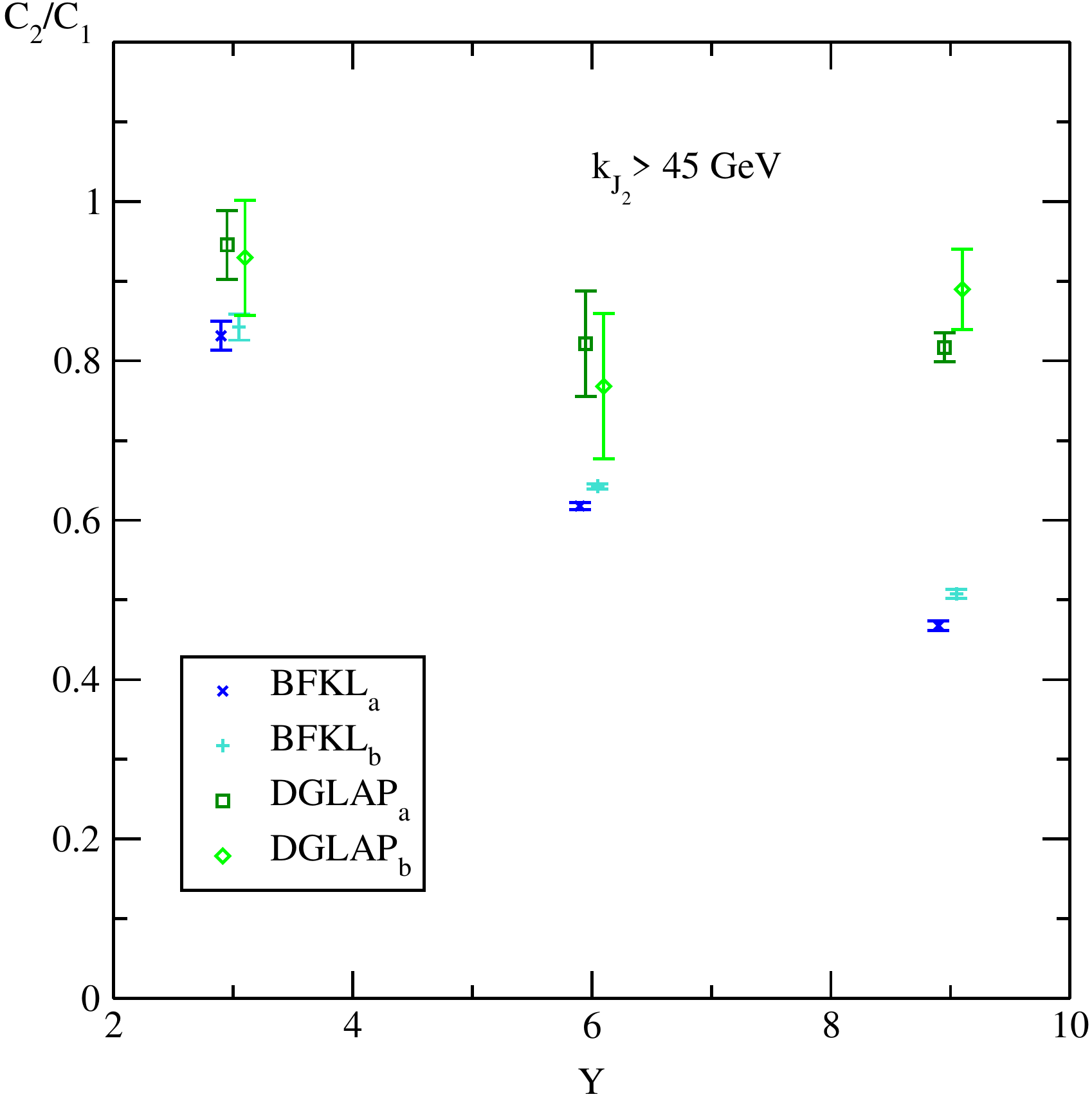}

   \includegraphics[scale=0.35]{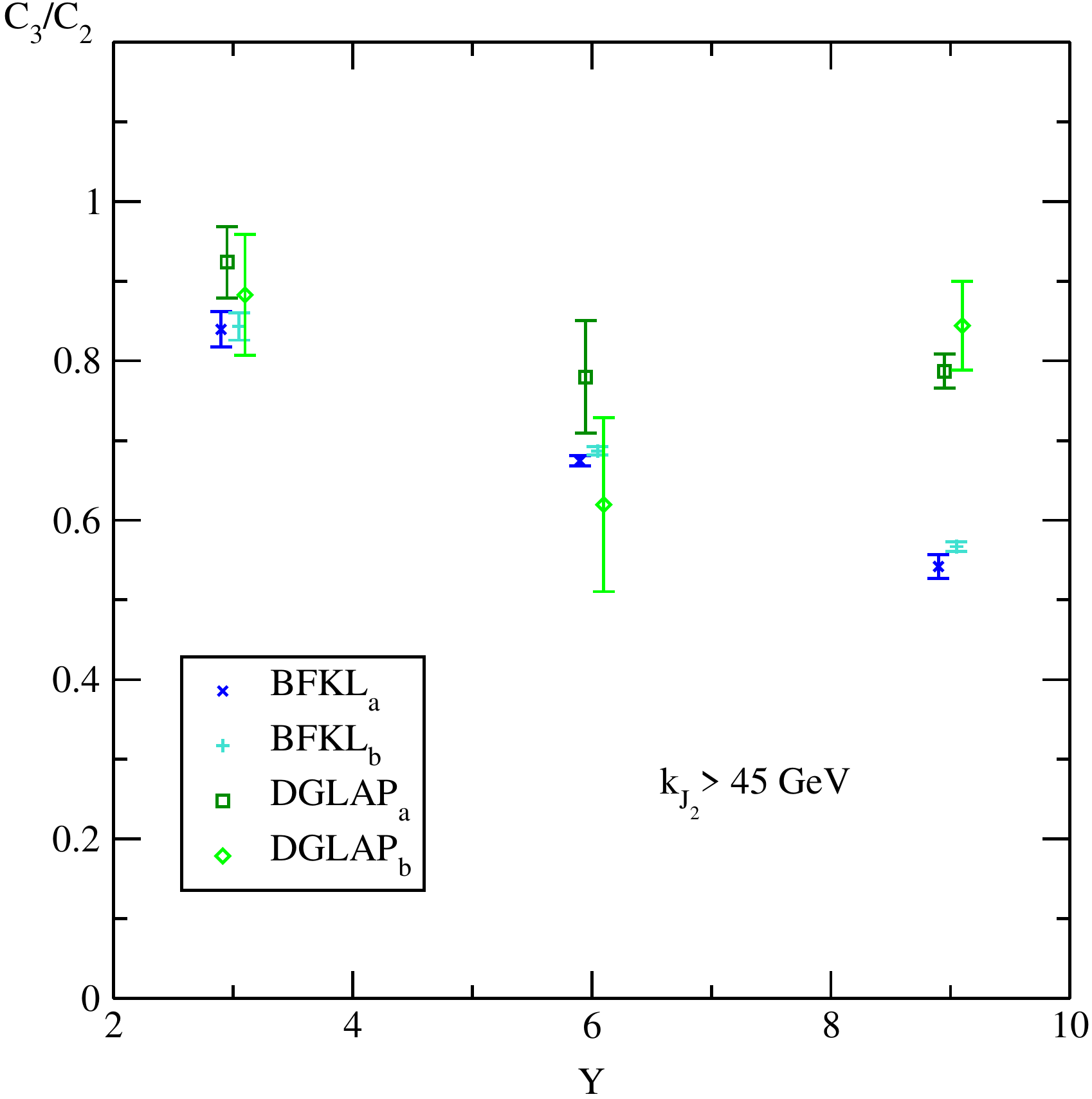}
\caption{$Y$-dependence of several ratios $C_m/C_n$ for $k_{J_1,\rm min}=35$ 
GeV and $k_{J_2,\rm min}=45$ GeV, for BFKL and DGLAP in the two variants of the 
BLM method (data points have been slightly shifted along the horizontal
axis for the sake of readability).}
\label{plot45}
\end{figure}

\begin{figure}[p]
\centering
   \includegraphics[scale=0.35]{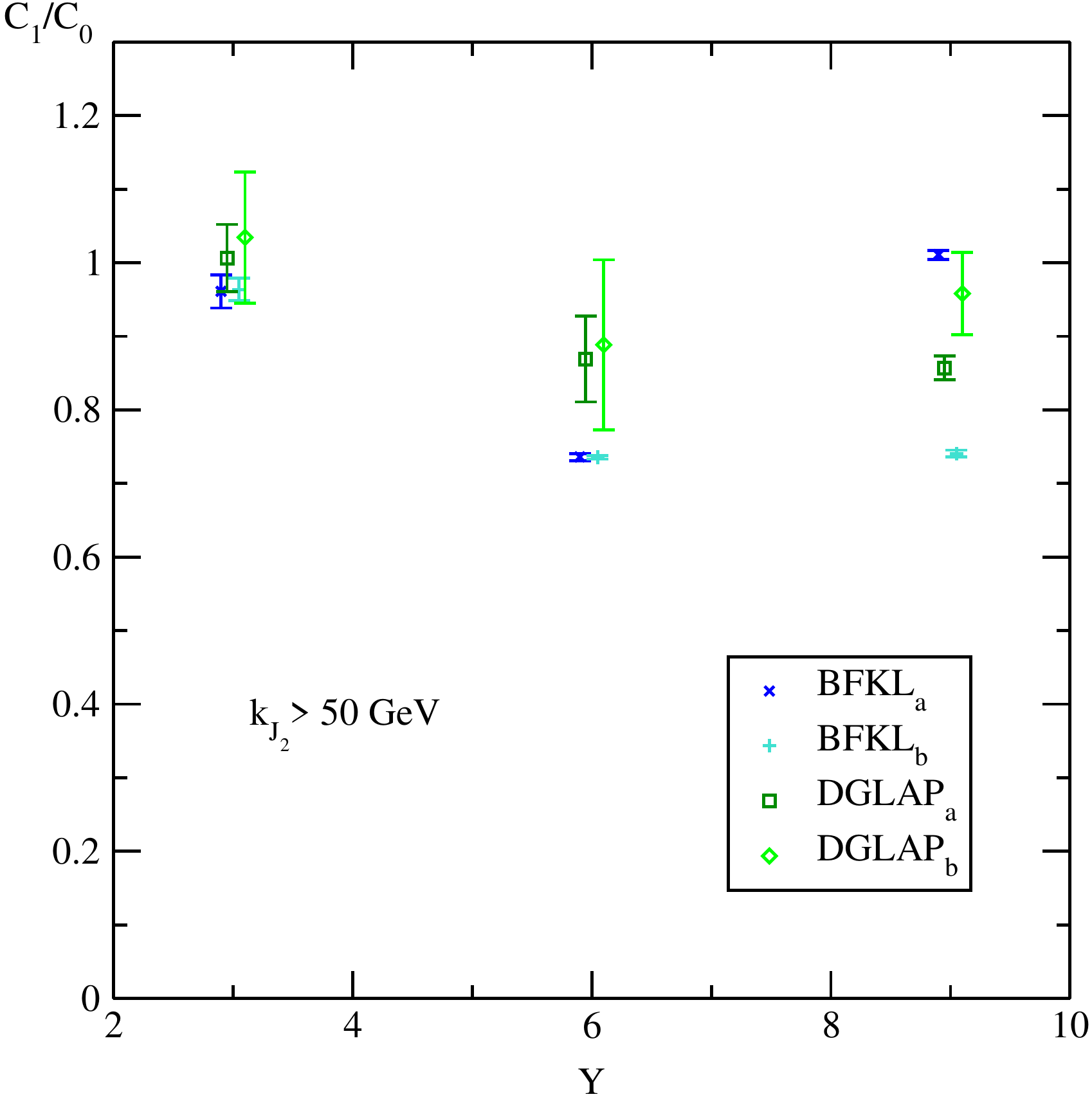}
   \includegraphics[scale=0.35]{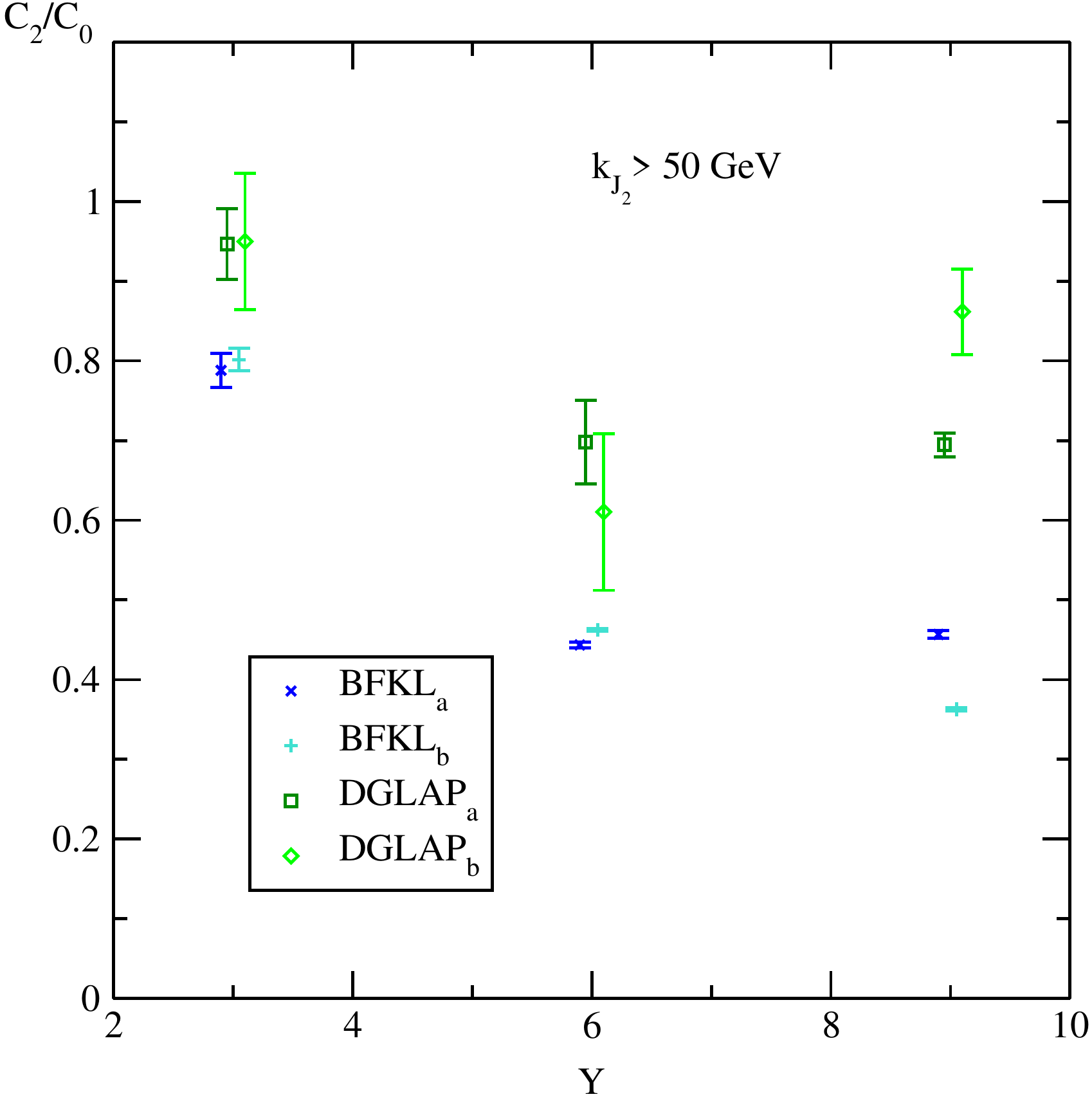}

   \includegraphics[scale=0.35]{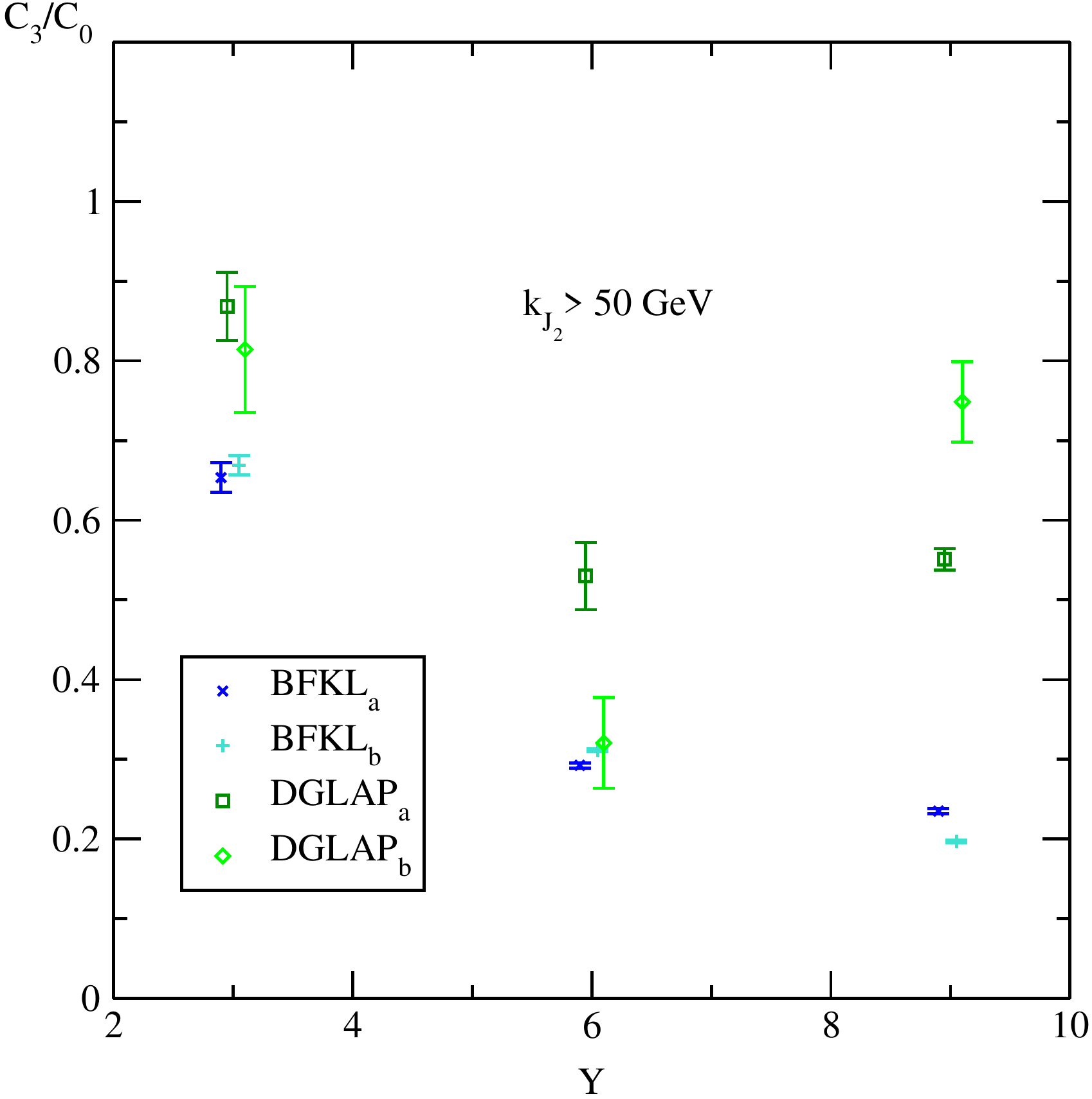}
   \includegraphics[scale=0.35]{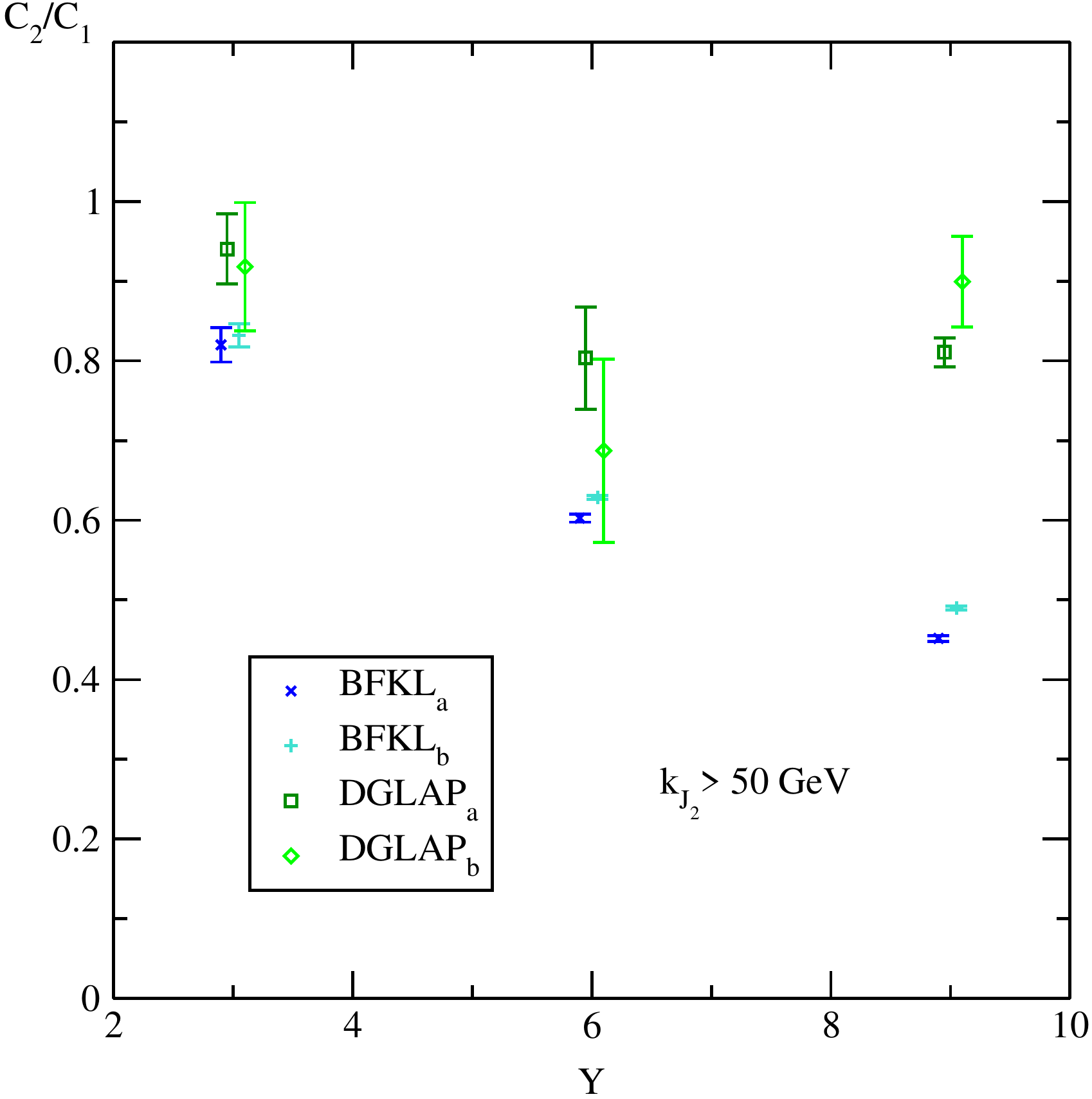}

   \includegraphics[scale=0.35]{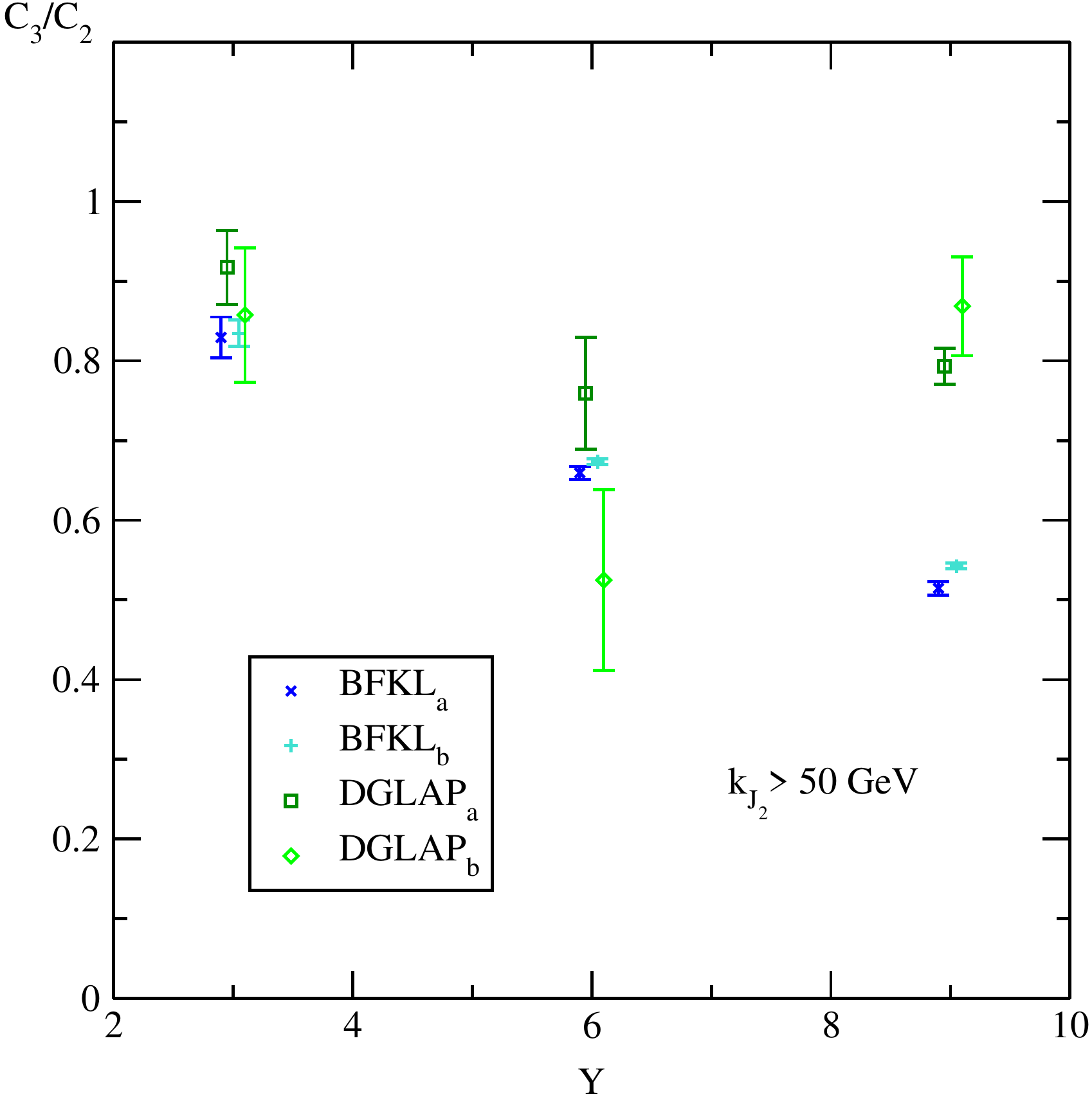}
\caption{$Y$-dependence of several ratios $C_m/C_n$ for $k_{J_1,\rm min}=35$ 
GeV and $k_{J_2,\rm min}=50$ GeV, for BFKL and DGLAP in the two variants of the 
BLM method (data points have been slightly shifted along the horizontal
axis for the sake of readability).}
\label{plots50}
\end{figure}

We summarize our results in Tables~\ref{tab:45} and~\ref{tab:50} and in 
Figs.~\ref{plot45} and~\ref{plots50}. We can clearly see that, at 
$Y=9$, BFKL and DGLAP, in both variants $(a)$ and $(b)$ of the
BLM setting, give quite different predictions for the all considered ratios 
except $C_1/C_0$; at $Y=6$ this happens in fewer cases, while at $Y=3$ 
BFKL and DGLAP cannot be distinguished with given uncertainties. This scenario
is similar in the two choices of the transverse momentum cuts.

\subsection{Used tools}
\label{num_tools}

All numerical calculations were implemented in \textsc{Fortran}, using
the corresponding interfaces for the MSTW 2008 PDFs~\cite{PDF}. 
Numerical integrations and the computation of the polygamma functions 
were performed using specific \textsc{CERN} program libraries~\cite{cernlib}.
Furthermore, we used slightly modified versions of the \cod{Chyp}~\cite{chyp} 
and \cod{Psi}~\cite{rpsi} routines in order to perform the calculation
of the Gauss hypergeometric function $_2F_1 $ and of the real part of the 
$\psi$ function, respectively. 

\subsection{Uncertainty estimation}
\label{num_uncertainty}

The are three main sources of uncertainty in our calculation:

\begin{itemize} 

\item The first source of uncertainty is the numerical 4-dimensional 
integration over the variables $|\vec k_{J_1}|$, $|\vec k_{J_2}|$, $y_{J_1}$ and
$\nu$ and was directly estimated by \cod{Dadmul} integration 
routine~\cite{cernlib}.

\item The second one is the one-dimensional integration over the longitudinal
momentum fraction $\zeta$ entering the expression for the NLO 
impact factors $c_{1,2}^{(1)}(n,\nu,|\vec k_{J_{1,2}}|, x_{J_{1,2}})$ given in
Eqs.~(36) and~(37) of Ref.~\cite{Caporale2013} and used in this work.
This integration was performed by using the \cod{WGauss} routine~\cite{cernlib}.
At first, we fixed the best value of the input accuracy parameter \cod{EPS}
by making comparisons between separate \textsc{Fortran} and 
\textsc{Mathematica} calculations of the impact factor. Then, we verified
that, under variations by factors of 10 or 1/10 of the \cod{EPS} parameter,
the $C_n^{\rm BFKL}$ and $C_n^{\rm DGLAP}$ coefficients change by less than 
1 permille.

\item The third one is related with the upper cutoff in the integrations
over $|\vec k_{J_1}|$, $|\vec k_{J_2}|$ and $\nu$. We fixed 
$k_{J_1 \rm max}=k_{J_2 \rm max} = 60$ GeV as in~\cite{Ducloue2013}, where it was
shown that the contribution to the integration from the omitted region
is negligible. Concerning the $\nu$-integration, we fixed the upper 
cutoff $\nu_{\rm max}=30$ for the calculation of the $C_n^{\rm BFKL}$ 
coefficients, after verifying that a larger value does not change the 
result in appreciable way. 

The $C_n^{\rm DGLAP}$ coefficients show a more pronounced sensitivity
to $\nu_{\rm max}$, due to the fact that the oscillations in the
integrand in Eqs.~(\ref{dglap}) and~(\ref{dglapb}) are not dumped by the 
exponential factor 
as in the BFKL expressions~(\ref{casea}) and~(\ref{caseb}). For the same reason,
the computational time of $C_n^{\rm DGLAP}$ is much larger than for 
$C_n^{\rm BFKL}$. We found that the best compromise was to set 
$\nu_{\rm max} = 50$. We checked in some sample cases, mostly at $Y=6$ and 9, 
that, putting $\nu_{\rm max}$ at 60, ratios $C_m/C_n$ change always less than
1\%, in spite of the fact that the single coefficients $C_n$ change
in a more pronounced way.

\end{itemize}

Of the three main sources of uncertainty, the first one is, by far, the most
significant, therefore the error bars of all data presented in this work
are just those given by the \cod{Dadmul} integration. We checked, however,
using some trial functions which mimic the behavior of the true integrands
involved in this work, that the error given by the \cod{Dadmul} integration
is a large overestimate of the true one. We are therefore confident that
our error estimation is quite conservative.



\section{Conclusions}
\label{conclusions}

In this paper we have considered the Mueller-Navelet jet production
process at LHC at the center-of-mass energy of 7~TeV and have compared
predictions for several azimuthal correlations and ratios between them,
both in full NLA BFKL approach and in fixed-order NLO DGLAP.

Differently from current experimental analyses of the same process, 
we have used {\it asymmetric} cuts for the transverse
momenta of the detected jets. In particular, taking one of the cuts 
at 35~GeV (as done by the CMS collaboration~\cite{CMS}) and the other at 
45~GeV or 50~GeV, we can clearly see that predictions from BFKL and DGLAP 
become separate for most azimuthal correlations and ratios between them,
this effect being more and more visible as the rapidity gap between the 
jets, $Y$, increases. In other words, in this kinematics the additional 
undetected parton radiation between the jets which is present in the resummed 
BFKL series, in comparison to just one undetected parton allowed by the NLO 
DGLAP approach, makes its difference and leads to more azimuthal angle 
decorrelation between the jets, in full agreement with the original proposal of 
Mueller and Navelet.

This result was not unexpected: the use of {\it symmetric} cuts for jet
transverse momenta maximizes the contribution of the Born term, which is present
for back-to-back jets only and is expected to be large, therefore making
less visible the effect of the BFKL resummation. This phenomenon
could be at the origin of the instabilities observed in the NLO fixed-order
calculations of~\cite{Andersen:2001kta,Fontannaz:2001nq}.

Another important benefit from the use of asymmetric cuts, pointed out 
in~\cite{Ducloue:2014koa}, is that the effect of violation of the 
energy-momentum conservation in the NLA is strongly suppressed with respect 
to what happens in the LLA.

In view of all these considerations, we strongly suggest experimental 
collaborations to consider also asymmetric cuts in jet transverse momenta in all
future analyses of Mueller-Navelet jet production process.

\vspace{1.0cm} \noindent
{\Large \bf Acknowledgments} \vspace{0.5cm}

We thank G.~Safronov for fruitful discussions. 
\\
\indent 
D.I. thanks the Dipartimento di Fisica dell'Universit\`a della Calabria and
the Isti\-tu\-to Na\-zio\-na\-le di Fisica Nucleare (INFN), Gruppo collegato 
di Cosenza, for warm hospitality and financial support.  The work of D.I. was 
also supported in part by the grant RFBR-15-02-05868-a. 
\\
\indent
B.M. thanks the Sobolev Institute of Mathematics of Novosibirsk for warm 
hospitality during the preparation of this paper. 
The work of B.M. was supported in part by the grant RFBR-13-02-90907 and by 
the European Commission, European Social
Fund and Calabria Region, that disclaim any liability for the use that can be
done of the information provided in this paper.




\begin{table}[p]
\centering
\caption{Ratios $C_n/C_m$ for $k_{J_1,\rm min}=35$ GeV and $k_{J_2,\rm min}=45$ GeV.}
\label{tab:45}
\begin{tabular}{c|c|llll}
\toprule
          & $Y$ & BFKL$_{(a)}$  & DGLAP$_{(a)}$  & BFKL$_{(b)}$  & DGLAP$_{(b)}$ \\
\midrule
          & 3.0 & 0.963(21)   & 1.003(44)   & 0.964(17)   & 1.021(78)   \\
$C_1/C_0$ & 6.0 & 0.7426(43)  & 0.884(61)   & 0.7433(30)  & 0.914(91)   \\
          & 9.0 & 0.897(15)   & 0.868(16)   & 0.714(10)   & 0.955(50)   \\
\midrule
          & 3.0 & 0.80(2)     & 0.948(43)   & 0.812(15)   & 0.949(75)   \\
$C_2/C_0$ & 6.0 & 0.4588(32)  & 0.726(56)   & 0.4777(26)  & 0.702(81)   \\
          & 9.0 & 0.4197(79)  & 0.710(15)   & 0.3627(50)  & 0.850(48)   \\
\midrule
          & 3.0 & 0.672(18)   & 0.876(41)   & 0.684(13)   & 0.838(70)   \\
$C_3/C_0$ & 6.0 & 0.3095(26)  & 0.566(45)   & 0.3282(21)  & 0.435(68)   \\
          & 9.0 & 0.2275(72)  & 0.558(13)   & 0.2057(29)  & 0.717(44)   \\
\midrule
          & 3.0 & 0.831(18)   & 0.945(43)   & 0.842(16)   & 0.929(72)   \\
$C_2/C_1$ & 6.0 & 0.6178(43)  & 0.821(66)   & 0.6427(34)  & 0.768(91)   \\
          & 9.0 & 0.4677(63)  & 0.817(18)   & 0.5079(56)  & 0.890(51)   \\
\midrule
          & 3.0 & 0.839(22)   & 0.924(45)   & 0.843(17)   & 0.883(76)   \\
$C_3/C_2$ & 6.0 & 0.6745(64)  & 0.780(71)   & 0.6869(52)  & 0.62(11)    \\
          & 9.0 & 0.542(15)   & 0.787(21)   & 0.5670(59)  & 0.844(56)   \\
\bottomrule
\end{tabular}
\end{table}

\begin{table}[p]
\centering
\caption{Ratios $C_n/C_m$ for $k_{J_1,\rm min}=35$ GeV and $k_{J_2,\rm min}=50$ GeV.}
\label{tab:50}
\begin{tabular}{c|c|llll}
\toprule
          & $Y$ & BFKL$_{(a)}$  & DGLAP$_{(a)}$ & BFKL$_{(b)}$  & DGLAP$_{(b)}$  \\
\midrule
          & 3.0 & 0.961(23)   & 1.006(46)   & 0.964(15)   & 1.034(89)   \\
$C_1/C_0$ & 6.0 & 0.7360(49)  & 0.869(58)   & 0.7357(25)  & 0.89(12)    \\
          & 9.0 & 1.0109(61)  & 0.857(16)   & 0.7406(46)  & 0.958(56)   \\
\midrule
          & 3.0 & 0.788(21)   & 0.946(44)   & 0.801(14)   & 0.950(85)   \\
$C_2/C_0$ & 6.0 & 0.4436(37)  & 0.698(53)   & 0.4626(19)  & 0.611(98)   \\
          & 9.0 & 0.4568(50)  & 0.695(15)   & 0.3629(23)  & 0.862(54)   \\
\midrule
          & 3.0 & 0.653(19)   & 0.868(43)   & 0.669(12)   & 0.814(79)   \\
$C_3/C_0$ & 6.0 & 0.2925(31)  & 0.530(42)   & 0.3115(15)  & 0.320(57)   \\
          & 9.0 & 0.2351(35)  & 0.551(17)   & 0.1969(17)  & 0.748(50)   \\
\midrule
          & 3.0 & 0.820(21)   & 0.940(44)   & 0.832(15)   & 0.918(81)   \\
$C_2/C_1$ & 6.0 & 0.6027(51)  & 0.803(64)   & 0.6288(26)  & 0.69(12)    \\
          & 9.0 & 0.4518(35)  & 0.811(18)   & 0.4900(24)  & 0.899(57)   \\
\midrule
          & 3.0 & 0.829(26)   & 0.917(46)   & 0.835(17)   & 0.857(85)   \\
$C_3/C_2$ & 6.0 & 0.6595(82)  & 0.759(70)   & 0.6733(36)  & 0.52(11)    \\
          & 9.0 & 0.5146(85)  & 0.793(23)   & 0.5426(38)  & 0.869(62)   \\
\bottomrule
\end{tabular}
\end{table}



\end{document}